\documentstyle[epsfig,prl,twocolumn,aps]{revtex}
\draft
\begin{document}
\wideabs
{\title{Phase diagram of an impurity in the spin-1/2 chain: two channel 
Kondo effect versus Curie law}
\author{Sebastian Eggert$^a$, David P.~Gustafsson$^a$, and Stefan Rommer$^b$}
\address{$^a$Institute of Theoretical Physics,
Chalmers University of Technology and G\"oteborg University,
S-412 96 G\"oteborg, Sweden
\\ $^b$Department of Physics and Astronomy, University of California,
 Irvine, California 92697
}
\date{Submitted: September 1, 2000.  Last change: \today}
\maketitle
\begin{abstract}
We consider a magnetic s=1/2 impurity in the antiferromagnetic spin 
chain as a function of two coupling parameters:  the symmetric 
coupling of the impurity to two sites in the chain 
$J_1$ and the coupling between the two
sites $J_2$.  By using field theory arguments and numerical calculations 
we can identify all possible fixed points and classify the renormalization 
flow between them, which leads to a non-trivial phase diagram. 
Depending on the detailed choice of the two (frustrating) coupling strengths, 
the stable phases correspond either to a decoupled spin with 
Curie law behavior
or to a non-Fermi liquid fixed point with a logarithmically diverging
impurity susceptibility as in the two channel Kondo effect. Our results 
resolve a controversy about the renormalization flow. 
\end{abstract}
\pacs{75.10.Jm, 75.20.Hr}}
\narrowtext

Impurities in low dimensional magnetic and electronic systems have
recently received a lot of attention in the context of
mesoscopic systems, quasi one-dimensional compounds and high temperature
superconductivity.
Much progress has been made in particular for one-dimensional systems, where 
quantum impurities are known to renormalize as the temperature is lowered,
so that the impurity can be described by an effective boundary condition
in the low temperature limit\cite{afflrev}.  
The temperature dependence of the impurity 
contributions to the specific heat and the susceptibility can also be 
predicted.   Very often the antiferromagnetic spin-1/2 chain with impurities
is taken as the prototype example for this renormalization 
behavior\cite{eggert,2CK,igar,zhang1,zhang2,zhang3,zhang4,rommer,clarke,frojdh,erik,qin,xxz,affleckedgefield,brunel}, which
captures the essential renormalization behavior of many other impurity systems. 

A number of different impurity configurations have been considered where 
typically one coupling parameter is varied.  The chain with one altered link 
is known to renormalize to an effective open boundary conditions\cite{eggert} in analogy 
with the renormalization of the conductivity in a quantum wire\cite{kane}.
However, two neighboring altered links renormalize to a periodic 
boundary condition\cite{eggert} in analogy with resonant tunneling in quantum 
wires\cite{kane2}.
This renormalization behavior is also equivalent to the two-channel Kondo 
effect\cite{afflrev,eggert,2CK}.  
Equally interesting is the coupling of an external impurity spin to
one or two sites in the chain.  An impurity spin $s$
coupled to one site in the chain gets screened with a resulting 
decoupled singlet of spin $s-1/2$ and an open chain with one site 
removed\cite{eggert,igar,zhang1,zhang2,zhang3,zhang4,rommer}.
An impurity spin coupled to two sites in the chain has first been considered
in Ref.~\onlinecite{clarke} where an equivalence with the two channel
Kondo effect was found on the basis of the field theory 
description\cite{clarke,frojdh}, which would
result in an ``overscreened'' impurity spin with a 
logarithmically diverging impurity susceptibility.
However, numerical simulation studies later questioned this result and predicted
a decoupled impurity spin as the stable boundary condition instead
with a Curie law susceptibility\cite{zhang3,zhang4}, which is an unresolved
controversy.

We now consider a two 
parameter impurity model. This allows us to consider the different 
impurity configurations above in a richer phase diagram 
with a greater number of possible fixed points.
Using field theory techniques and advanced numerical results we are
able to map out the phase separation line between an overscreened 
and a decoupled impurity spin exactly, which not only resolves the 
controversy, but also shows a more complex renormalization behavior.

\begin{figure}
\begin{center}
\mbox{\epsfig{file=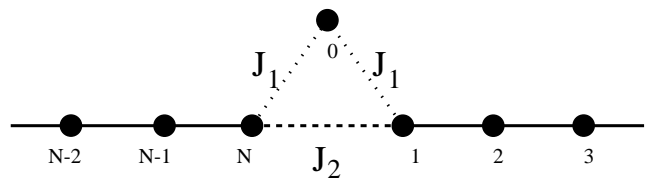,width=3.35in,angle=0}}
\end{center}
\caption{Impurity model with two parameters $J_1$ and $J_2$.}
\label{impmodel}
\end{figure}

The Hamiltonian we consider here describes a Heisenberg chain where two
neighboring sites are coupled to an impurity 
spin-1/2 with strength $J_1$. The two sites are also coupled
to each other with strength $J_2$
as shown in Fig.~\ref{impmodel}
\begin{equation}
H = J \sum_{i =  1}^{N-1} 
\vec{S}_i \cdot \vec{S}_{i+1} + J_1 \vec{S}_{\rm 0} \cdot 
\left(\vec{S}_{N} + \vec{S}_1 \right) + J_2 \vec{S}_{N} \cdot  \vec{S}_1.
\label{ham}
\end{equation}
In the low temperature limit this system is known to be well described by 
by a level 1 Wess-Zumino-Witten (WZW) model with a marginal irrelevant 
operator\cite{affleck}.
A spin operator at position $x$ in the chain can be expressed in terms of 
the current operators $\vec{J}$ and the WZW field $g$
\begin{equation}
\vec{S}(x) \approx \vec{J}_L +  \vec{J}_R + (-1)^x {\rm const.\ } 
{\rm tr} \vec{\sigma} g. \label{spin}
\end{equation} 

\begin{figure}
\begin{center}
\mbox{\epsfig{file=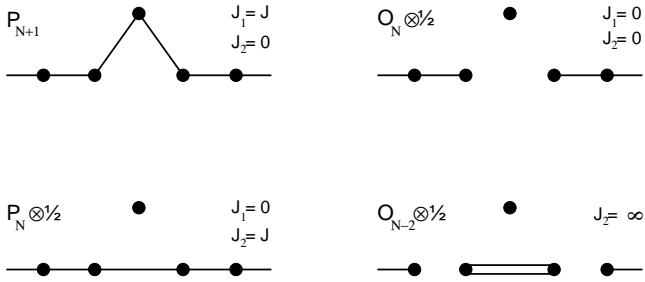,width=3.35in,angle=0}}
\end{center}
\caption{The four possible fixed points.}
\label{fixedpoints}
\end{figure}

General values of $J_1$ and $J_2$ break the conformal 
invariance of the effective field theory and introduce perturbing 
operators in the field theory Hamiltonian,
which in turn can be classified by their renormalization behavior.  
As the temperature is lowered the system always must approach a 
fixed point that restores conformal invariance\cite{afflrev}.
For the spin-1/2 chain in Eq.~(\ref{ham}) the possible fixed points are
always given in terms of simple boundary conditions on the spin chain
(open or periodic).
In addition there may be completely decoupled impurity spins present.
The possible fixed points for this system are shown in Fig.~\ref{fixedpoints}.
\begin{itemize}
\item $J_1 = J, \ J_2 =  0$. A periodic chain with $N+1$ sites and no impurity 
spin. The impurity spin has been absorbed. We denote this fixed point 
by $\rm \mathbf P_{N+1}$.
\item $J_1 = 0, \ J_2 =  0$. An open chain with $N$ sites and a decoupled s=1/2 
impurity spin.  We denote this fixed point by  $\mathbf O_N \otimes \case{1}{2}$.
\item $J_1 = 0, \ J_2 =  J$. A periodic chain with $N$ sites and a decoupled 
s=1/2 impurity spin, denoted by $\rm \mathbf P_N \otimes \case{1}{2}$.
\item $J_1 = 0, \  J_2 =  \infty$. 
An open chain with $N-2$ sites and a decoupled
impurity s=1/2.  The two end sites at $1$ and $N$ have locked into a
 singlet state.  We denote this fixed point 
by $\rm \mathbf O_{N-2} \otimes \case{1}{2}$.
\end{itemize}

The renormalization behavior in the vicinity of each 
of those fixed points is determined by the operator content
of the impurity Hamiltonian according to the perturbation of $J_1$ and
$J_2$ around the corresponding fixed point values.
Typically the operators with the lowest scaling dimensions (leading operators) 
can be determined by either a 
straight-forward symmetry analysis or by direct application of Eq.~(\ref{spin}).
Local operators become relevant when their scaling dimensions is less than
one $d < 1$. 

\begin{figure}
\begin{center}
\mbox{\epsfig{file=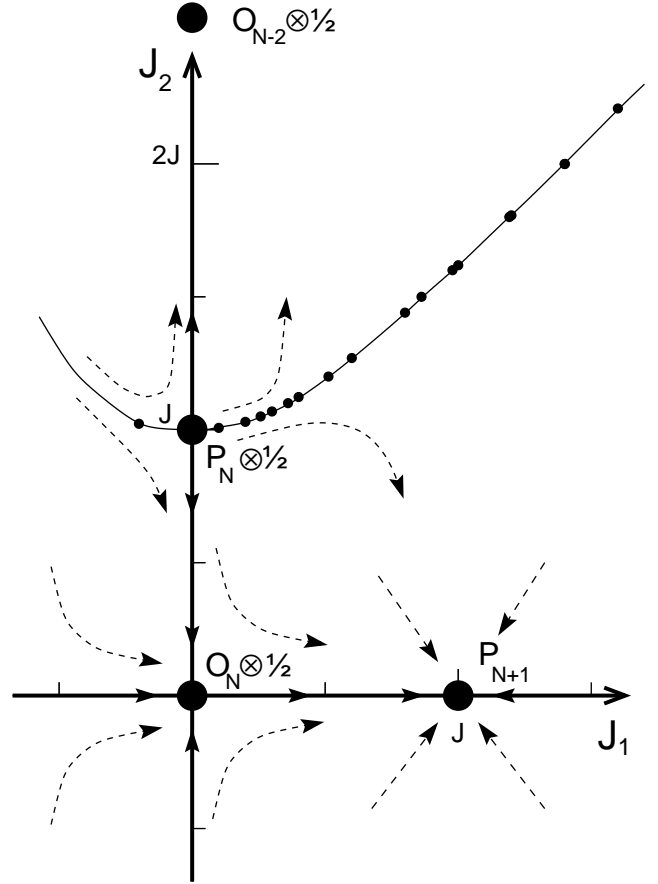,width=3.35in,angle=0}}
\end{center}
\caption{The full phase diagram of the $J_1-J_2$ impurity model.
The possible fixed points are indicated by the thick solid dots.
The points on the phase separation line have been estimated numerically 
by analyzing $\langle \vec{S}_1 \cdot \vec{S}_N \rangle$ as in 
Fig.~\ref{corr} for 132 different values for $J_1$ and $J_2$.}
\label{phasediagram}
\end{figure}

In the vicinity of the periodic fixed point $\rm \mathbf P_{N+1}$ the 
leading irrelevant operator is given by $\partial_x {\rm tr} g$
because site parity symmetry does not allow any more relevant
operators\cite{eggert}.  We therefore immediately conclude that
this fixed point is stable in all directions in the $J_1-J_2$
phase diagram, i.e.~small perturbations on both 
$J_1$ and $J_2$ are irrelevant as indicated in the phase 
diagram Fig~\ref{phasediagram}.


In the vicinity of the the fixed point $\rm \mathbf O_{N} \otimes \case{1}{2}$ 
the fields in the perturbing Hamiltonian are given by boundary operators.
This stems from the analytic continuation of the left moving fields in terms
of right moving fields in order to restore the conformal invariance at 
an open boundary.  The leading operator 
$(\vec{J}_L(0) + \vec{J}_L(N))\cdot \vec{S}_{\rm imp}$ of
scaling dimension $d=1$  
is created by the coupling $J_1$ to the impurity spin from 
the open ends.
This operator is marginally relevant for 
antiferromagnetic sign and marginally irrelevant for ferromagnetic 
sign analogous to the Kondo effect\cite{eggert}.  The coupling
between the end spins $J_2$ can only produce the 
irrelevant operators\cite{eggert} $\vec{J}_L^2(0),  \vec{J}_L^2(N)$ and
$\vec{J}_L(0) \cdot \vec{J}_L(N)$ with scaling dimension $d=2$.
Hence the coupling
$J_2$ is always irrelevant in the vicinity of $\rm O_{N} \otimes \case{1}{2}$.
We therefore conclude that the fixed point 
$\rm O_{N} \otimes \case{1}{2}$ is stable for $J_1 \leq 0$, but for $J_1 > 0$
the effective coupling to the impurity spin increases under renormalization 
and flows towards the stable fixed point $\rm P_{\rm N+1}$, 
corresponding to a ``healing'' of the chain\cite{eggert}
as indicated in the phase diagram Fig~\ref{phasediagram}.  
This effect is in fact 
completely analogous to the renormalization flow and overscreening 
of the spin excitations
in the two-channel Kondo problem\cite{afflrev,eggert,2CK}, resulting
in a logarithmically diverging impurity susceptibility.

%

Near the fixed point
$\rm \mathbf O_{N-2} \otimes \case{1}{2}$ the impurity spin is separated by a 
locked singlet in the limit $J_2 \to \infty$ which is
effectively decoupled from the rest of the chain (unless we
introduce additional coupling constants).  Therefore the fixed point 
$\rm O_{N-2} \otimes \case{1}{2}$ appears to be stable in all directions in 
our $J_1-J_2$ phase diagram in Fig~\ref{phasediagram}.

A more interesting scenario can be found near the fixed point
$\rm \mathbf  P_N \otimes \case{1}{2}$ since a small coupling $J_1$ of either 
sign is frustrating. Here the most relevant 
operator ${\rm tr} g$ corresponds to a slight modification of one link in the
chain $J_2$. If the impurity spin is absent this operator is responsible for
the breaking of the chain\cite{eggert}, but the frustrating
coupling $J_1$ must now also be considered.
In particular, a small coupling to the impurity spin $J_1$ produces the
operator $(\vec{J}_L + \vec{J}_R)\cdot \vec{S}_{\rm imp}$
which is marginally relevant/irrelevant for antiferromagnetic/ferromagnetic
sign respectively.  The irrelevant operator 
$\partial_x {\rm tr} \vec{\sigma}g \cdot \vec{S}_{\rm imp}$ of dimension $d=3/2$
is also created by $J_1$.  In summary we may express the impurity 
Hamiltonian at the fixed point $\rm P_N \otimes \case{1}{2}$ as
\begin{equation}
H_{\rm imp}  =
\gamma_1 {\rm tr} g +
\gamma_2 (\vec{J}_L + \vec{J}_R)\cdot \vec{S}_{\rm imp} +
\gamma_3 \partial_x {\rm tr} \vec{\sigma}g \cdot \vec{S}_{\rm imp},
\label{h-imp-3}
\end{equation}
where to lowest order
\begin{equation}
\gamma_1 \propto (J_2-J), \ \ \ 
\gamma_2 \propto J_1, \ \ \  \gamma_3 \propto J_1. \label{couplconst}
\end{equation}

The renormalization group equations can then be determined 
by using the
operator product expansion.  We find that
\begin{eqnarray}
\dot \gamma_1 & = & \case{1}{2}\gamma_1 - \case{3}{2} \gamma_2 \gamma_3 \nonumber \\
\dot \gamma_2 & = & \gamma_2^2 - \case{3}{4} \gamma_3^2 
\label{reneqns}\\
\dot \gamma_3 & = & -\case{1}{2} \gamma_3 + 2 \gamma_2 \gamma_3.
\nonumber 
\end{eqnarray}
where the dot indicates the derivative in respect to the logarithm of the 
cutoff.
The renormalization flow is dominated by the relevant 
coupling constant $\gamma_1$, so that it is crucial to determine 
if  $\dot \gamma_1$ is initially negative or
positive.  Interestingly, the irrelevant coupling constant 
$\gamma_3$ plays an important role in defining the corresponding 
phase separation line along $\gamma_1  = 3 \gamma_2 \gamma_3$, which together
with Eq.~(\ref{couplconst}) gives a parabolic shape in the $J_1-J_2$ phase
diagram $J_2-J \propto J_1^2$ as indicated in Fig.~\ref{phasediagram}. 
Indeed if the irrelevant coupling constant $\gamma_3$ is neglected, 
one might erroneously come to the conclusion that 
the relevant backscattering $\gamma_1$ is exactly zero 
along the line $J_2 = J$ resulting in just a marginal renormalization flow
as postulated in Refs.~\onlinecite{clarke,frojdh} which is
not the case.

If $\dot\gamma_1> 0$ the effective coupling $J_2$ increases
quickly so that the two end spins lock into a singlet independent of the 
value of $J_1$.  Therefore, the stable fixed point is 
$\rm O_{N-2} \otimes \case{1}{2}$ above the phase separation line.
If, however, $\dot\gamma_1< 0$ the effective coupling $J_2$
decreases towards zero and the stable fixed point now depends on the 
marginal coupling $J_1$.  For $J_1 \leq 0$ the stable fixed point is 
$\rm O_{N} \otimes \case{1}{2}$ while for $J_1 > 0$ the ``healing'' process 
towards the fixed point $\rm P_{N+1}$ takes place again.  
The leading irrelevant operators close to $\rm P_{N+1}$ will 
again produce a logarithmically divergent impurity susceptibility\cite{afflrev},
but the renormalization flow 
is not completely analogous to the two channel Kondo effect since
the operator content near the unstable fixed point is quite different.
In particular the relevant operator ${\rm tr} g$ is completely absent in the
ordinary Kondo problem.  The impurity susceptibility in the phases characterized
by the fixed points $\rm O_{N} \otimes \case{1}{2}$ and
$\rm O_{N-2} \otimes \case{1}{2}$ is a Curie law behavior as $T\to 0$
from the decoupled impurity spin degrees of freedom.

We have now determined the basic shape of the phase diagram, but 
to quantitatively map out the phase separation line it is necessary to resort
to numerical methods.  We therefore choose the transfer matrix 
renormalization group method for impurities, which can determine 
local correlation functions directly in the thermodynamic 
limit\cite{rommer}.
One obvious order parameter is the correlation between the 
two end spins which takes well-defined values at each fixed point
in the zero temperature limit
\begin{eqnarray}
&{\rm P_{N+1} }   \phantom{MMMMM}  &\langle \vec{S}_1 \cdot \vec{S}_N \rangle =
\case{1}{4} - 4 \ln 2 + \case{9}{4} \zeta(3)  \nonumber \\
& {\rm O_N \otimes \case{1}{2}} \phantom{MMMMM}  &\langle \vec{S}_1 \cdot \vec{S}_N \rangle = 0 \nonumber \\
& {\rm P_N \otimes \case{1}{2}}  \phantom{MMMMM} & \langle \vec{S}_1 \cdot \vec{S}_N \rangle =
\case{1}{4} -  \ln 2   \nonumber \\
& {\rm O_{N-2} \otimes \case{1}{2}} \phantom{MMMMM} & \langle \vec{S}_1 \cdot \vec{S}_N \rangle 
=-\case{3}{4} , \label{correlations} 
\end{eqnarray}
where $\zeta$ is the Riemann Zeta function.
The values at the periodic chain fixed points correspond 
to the first and second nearest neighbor 
correlations in the ordinary Heisenberg chain\cite{taka}, while the open
chain fixed point values
correspond to two uncorrelated spins and a spin singlet, respectively.
Numerical simulations can never reach zero
temperature in the thermodynamic limit, but it is well possible to 
determine if the correlation between the spins renormalizes towards
larger or smaller values as the temperature is lowered and thereby determine
the stable fixed point.  

In particular we postulated that above the phase-transition line in 
Fig.~\ref{phasediagram} the two end spins effectively lock into a singlet
so that the correlation should {\it decrease} with decreasing temperature.
Below the phase transition line, however, the two spins will only
be weakly correlated at $\rm O_N \otimes \case{1}{2}$ or ferromagnetically
correlated at $\rm P_{N+1}$ so that the correlation should {\it increase}
with decreasing temperature.

\begin{figure}
\begin{center}
\mbox{\epsfig{file=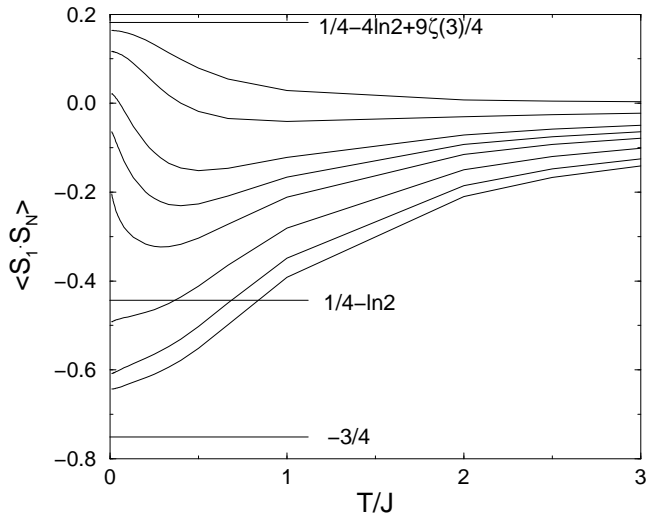,width=3.35in,angle=0}}
\end{center}
\caption{The correlation $\langle \vec{S}_1 \cdot \vec{S}_N \rangle$ 
as a function of $T$ for 
$J_1 = 0.8 J$ and $J_2 \ = \ 0, \: 0.4, \: 0.8,\: 1,\: 1.2,\: 1.5,\: 1.8,\: 2 J$ 
from above.  By analyzing the slope or absolute
value compared to $\case{1}{4} - \ln 2$ the phase transition can be 
estimated to be at $J_2 \approx 1.44 J$ in this case.}
\label{corr}
\end{figure}

This behavior is shown in Fig.~\ref{corr} for $J_1 = 0.8J$ and different
values of $J_2$.  As $T\to 0$ the correlation clearly decreases for larger values
of $J_2 \agt 1.5 J$  but increases for smaller values $J_2 \alt 1.2 J$ so
that the phase transition occurs somewhere between
those two values.
We can identify the critical value of $J_2$ more closely  
by considering the slope
and absolute value of $\langle \vec{S}_1 \cdot \vec{S}_N \rangle$ as
$T\to 0$.  The critical value can  be defined as the point where
the slope is zero or the absolute value is $1/4-\ln2$, which gives 
approximately the same result of $J_2 \approx 1.44J$.
We have made a similar analysis for many different points in the
$J_1-J_2$ parameter space and thereby mapped out the exact 
numerical location
of the phase separation line as shown in Fig.~\ref{phasediagram}.
As $J_1, J_2 \to \infty$ the phase transition occurs at $J_1 = J_2$
which can be determined by analyzing the ground state of the
three coupled spins $\vec{S}_N,\ \vec{S}_0,\  \vec{S}_1$.

We now wish to relate our results to previous numerical studies 
of this impurity model by Zhang et al\cite{zhang2,zhang3,zhang4}, 
where three  points 
in the phase diagram were analyzed at $(J_1 = 0.4 J, \ J_2 = J)$, 
$(J_1 = 0.4 J, \ J_2 = 0)$, and
$(J_1 = 1.5 J, \ J_2 = J)$, which led to a controversy with the predictions
in Refs.~\onlinecite{clarke,frojdh}.  In particular, the numerical 
studies found no 
logarithmically diverging impurity susceptibility at $(J_1 = 0.4J,\  J_2 = J)$. 
As mentioned above the analytic studies in 
Refs.~\onlinecite{clarke,frojdh} failed to 
consider the leading relevant operator ${\rm tr} g$ at the unstable 
fixed point $\rm P_{N} \otimes \case{1}{2}$,  but
the logarithmic impurity susceptibility is 
created by the  leading irrelevant operator $\partial_x  {\rm tr g}$
near the {\it stable} fixed point $\rm P_{N+1}$\cite{afflrev},
so that we also 
postulate a logarithmic divergence below some crossover temperature $T_K$.  
However, looking at the phase diagram in Fig.~\ref{phasediagram} we see
that the point $(J_1 = 0.4 J, \ J_2 = J)$ lies very close
to the phase transition line, so that the crossover temperature $T_K$ must be
extremely small and cannot be reached by any numerical method today, which 
explains the controversy.
On the other hand the points $(J_1 = 1.5, \ J_2 = J)$ 
and $(J_1 = 0.4, \ J_2 = 0)$ are further from the phase separation and
the crossover temperature is larger. Therefore, the
 logarithmic divergence can indeed 
be observed\cite{2CK,zhang2,zhang4}.  In Ref.~\onlinecite{zhang4} the 
end spin correlation $\langle \vec{S}_1 \cdot \vec{S}_N \rangle$
was also determined, which is consistent with our results, but with a different 
interpretation. Because the overall value of 
$\langle \vec{S}_1 \cdot \vec{S}_N \rangle$ is negative it was 
concluded in that work that the point $(J_1 = 0.4J,\  J_2 = J)$ produces
a Curie law susceptibility from frustration, i.e.~an effectively  decoupled 
impurity spin. However, the temperature dependence clearly shows a 
renormalization towards an
increasing correlation as $T\to 0$, so that 
$\rm P_{N+1}$ is the correct stable fixed point with an absorbed
impurity spin and a logarithmically diverging susceptibility below an
extremely small crossover temperature $T_K$.
Interestingly, there is an integrable spin-1 chain model 
with a spin-1/2 impurity which can also reproduce 
this two-channel Kondo renormalization behavior 
near the fixed point $\rm P_{N+1}$\cite{schlottmann}.  However,
the interesting boundary fixed points that were recently discovered
in this integrable model\cite{dai} have no analogous expression in our
parameter space.

In conclusion we have analyzed a complex impurity model in the spin-1/2 chain
which shows a rich phase diagram.   We were able to identify three different
stable phases that are attracted to well-defined fixed points under
renormalization.  The three phases are separated by the unstable 
fixed point $\rm P_{N} \otimes \case{1}{2}$ which 
shows a high sensitivity to the detailed choice 
of couplings due to frustration.  Using field theory arguments and 
numerical techniques we were able to map out the phase diagram exactly and
also resolve a controversy in the literature\cite{zhang3,zhang4,clarke,frojdh}.
Interestingly, an irrelevant operator plays a crucial rule in determining
the shape of the phase separation line.  It would be interesting to see
if the current findings can be generalized to other systems where 
a frustrated state may separate more stable phases (e.g. in
a ladder system with a zigzag geometry).  In general, a frustrated
state should always be unstable since it has been postulated that 
renormalization typically occurs towards a state with a lower 
ground state degeneracy\cite{g-theorem}. The system we considered here 
is a typical example of this behavior.

\begin{acknowledgements}
This research was supported in part by the
Swedish Natural Science Research Council (NFR)
and the Swedish Foundation for International
Cooperation in Research and Higher Education (STINT).  
\end{acknowledgements}


%
%
%

\end{document}